\documentclass[
    ,final            
  ]
  {aipproc}

\layoutstyle{6x9}

\begin{document}

\title{$\pi^{0} $ Transverse Single-Spin Asymmetries at $\eta=4.1$ 
in $p+p$ Collisions at $\sqrt{s}=200$ GeV}



\author{J.~L. Drachenberg$^1$, for the STAR Collaboration}{
  address={$^1$~Cyclotron Institute, Texas A\&M University, College Station, TX  77843}
}

\begin{abstract}
Large transverse single-spin asymmetries, $A_{N}$, have been observed in forward hadron production 
at RHIC. STAR has reported precision measurements of the $x_{F}$-$p_{T}$ dependence of $A_{N}$ 
for forward $\pi^{0}$ production. Contrary to expectation, the asymmetry does not fall with $p_{T}$. 
However, questions remain about the trend at the lower end of the studied range, $p_{T}\sim1$ GeV/c. 
Analysis of data from the 2008 RHIC run has extended the $\pi^{0}$ data in the $x_{F} > 0.4$, low 
$p_{T}$ region. Preliminary results are reported.
\end{abstract}

\maketitle


\section{Introduction}

In 2006, STAR measured transverse single-spin asymmetries,
$A_{N}$, for forward rapidity $\pi^{0}$ production in $p+p$ collisions at $\sqrt{s}=200$ GeV to 
$p_{T}\sim4$ GeV \cite{Abelev:2008}. Results showed large $A_{N}$ at high-$x_{F}$ as 
seen in previous experiments at lower center-of-mass energies, such as Fermilab E704 
\cite{Adams:1991-261,Adams:1991-264}. In contrast to the E704 kinematics, pQCD predictions provide 
a reasonable description of the observed cross-sections for forward $\pi^0$ production at 
$\sqrt{s}=200$ GeV \cite{Adams:2006}. Model calculations based on the Sivers \cite{D'Alesio:2004} and 
Collins \cite{Yuan:2008} effects provide fair descriptions of the $x_{F}$-dependence of the measured 
asymmetries. The Sivers model relates the asymmetry to correlations involving initial-state 
parton transverse momentum \cite{Sivers:1990} while the Collins model relates the asymmetry to final-state transverse momentum dependent fragmentation \cite{Collins:1993}. However, the observed 
asymmetries do not provide an indication of decreasing $A_{N}$ with increasing $p_{T}$, as predicted 
by theory \cite{D'Alesio:2004,Yuan:2008}. Isolating the origin or, in the case of a combination of effects, disentangling the origins of the 
observed asymmetries would provide crucial information about proton spin structure.

The previous STAR measurements \cite{Abelev:2008} provide a hint of an initial decrease of 
$A_{N}$ with $p_{T}$ for $x_F > 0.4$. The data obtained in this low-$p_{T}$, high-$x_{F}$ region were 
recorded during 2003 and 2005, and are limited by low statistics and low polarization. For the 2008 
RHIC run, STAR extended measurements to low-$p_{T}$, high-$x_{F}$ kinematics to further investigate 
this region with better statistics and polarization.

\begin{figure}[t]
  \scalebox{1}
{    \includegraphics*[scale=.58]{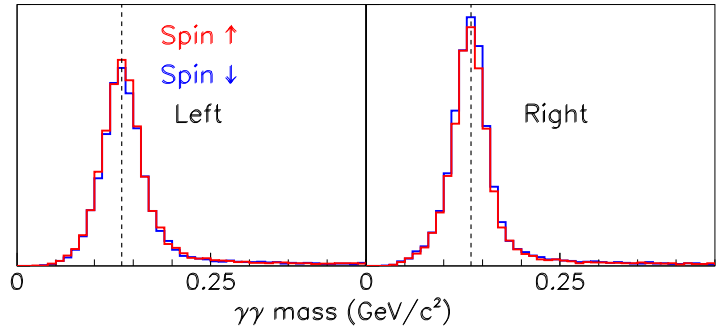}
}
    \caption{Spin-sorted di-photon mass peaks. The dashed lines mark the known mass value of the 
    $\pi^{0}$. 
    \label{fig:spinsort}}
\end{figure}

\section{Forward Pion Detector}

To obtain the measurements for the 2008 run, data were taken with the STAR Forward Pion Detector 
(FPD). The FPD is an electromagnetic calorimeter, consisting of two modules, left and right of the 
beamline on the east side of STAR. Each module consists of a $7\times7$ array of PbGl calorimeters 
positioned 809 cm from the interaction point \cite{Bland:2006}. Each PbGl calorimeter cell measures 
$3.8\times3.8\times45$ cm$^{3}$. To access the desired low-$p_{T}$, high-$x_{F}$ kinematics, the 
modules were positioned at $<\eta>\sim-4.1$.

\section{Event Cuts and Systematics}

%
%
Data were taken from $p_{\uparrow}+p\rightarrow\pi^{0}+X$ yielding
$L_{int}\sim0.5$ pb$^{-1}$ with beam polarization, $P_{beam}\sim0.44$. Events were
required to pass a hardware trigger that required a nominal summed energy of 
25 GeV in a single FPD module. Software cuts were applied for di-photon events, requiring energy 
sharing $z_{\gamma\gamma}<0.7$ and $0.07$ GeV/c$^{2}$ 
$<m_{\gamma\gamma}<0.3$ GeV/c$^{2}$, and a software total energy threshold of $E_{total}>25$ 
GeV was applied. The events were required to fall within a fiducial volume of one-half the width of a  
PbGl cell from the edge of each module. Calibration procedures involved analyzing raw ADC's 
channel-by-channel and run-by-run for pedestal shifts. For gain calibration, $\pi^{0}$ mass peaks 
were reconstructed channel-by-channel and corrected in an iterative process until convergence 
to the known value was achieved. Energy-dependent corrections were evaluated by fitting the 
$\pi^{0}$ peak in energy bins, then obtaining a linear fit of the ratio of the known value over the fitted 
centroids vs. energy. Finally, run-by-run summed-mass values are used to correct for run-dependent 
gain shifts. The result is agreement between the di-photon peaks in each module and the known value of the $\pi^{0}$ for polarization spin-up or spin-down, Fig.\@ \ref{fig:spinsort}.

Systmatic errors are dominated by gain calibration issues and yields
under the pion mass peak. Gain calibration issues arise because of
limited statistics in the PbGl cells that are farthest from the beam in each module.
To correct for yields under the pion peak, varying mass cuts were studied. The total systematic 
uncertainty was obtained by combining the effects in quadrature.

\section{Run-8 Asymmetries}

\begin{figure}
  \centering
  \includegraphics*[scale=0.35]{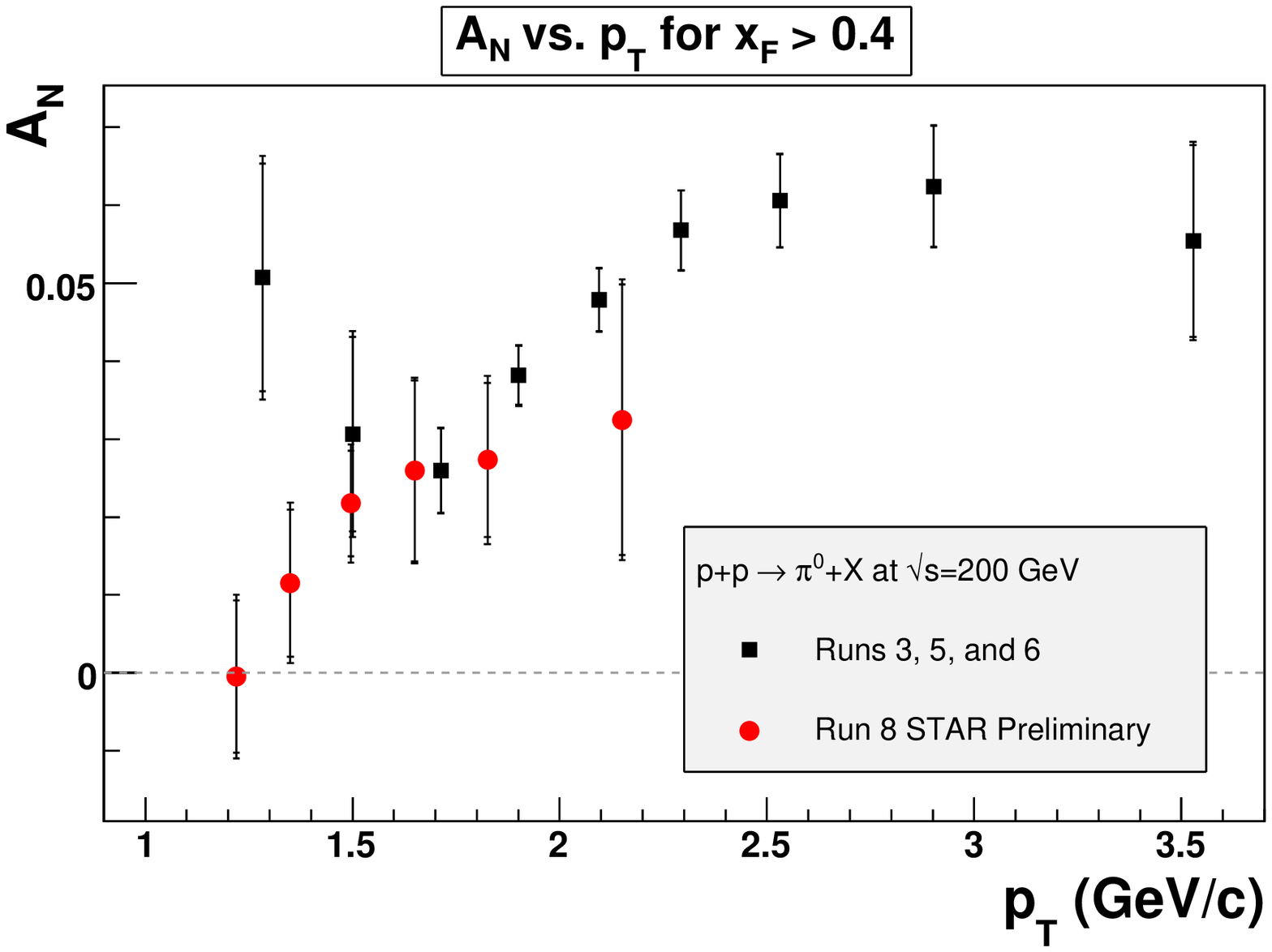}
  \includegraphics*[scale=0.35]{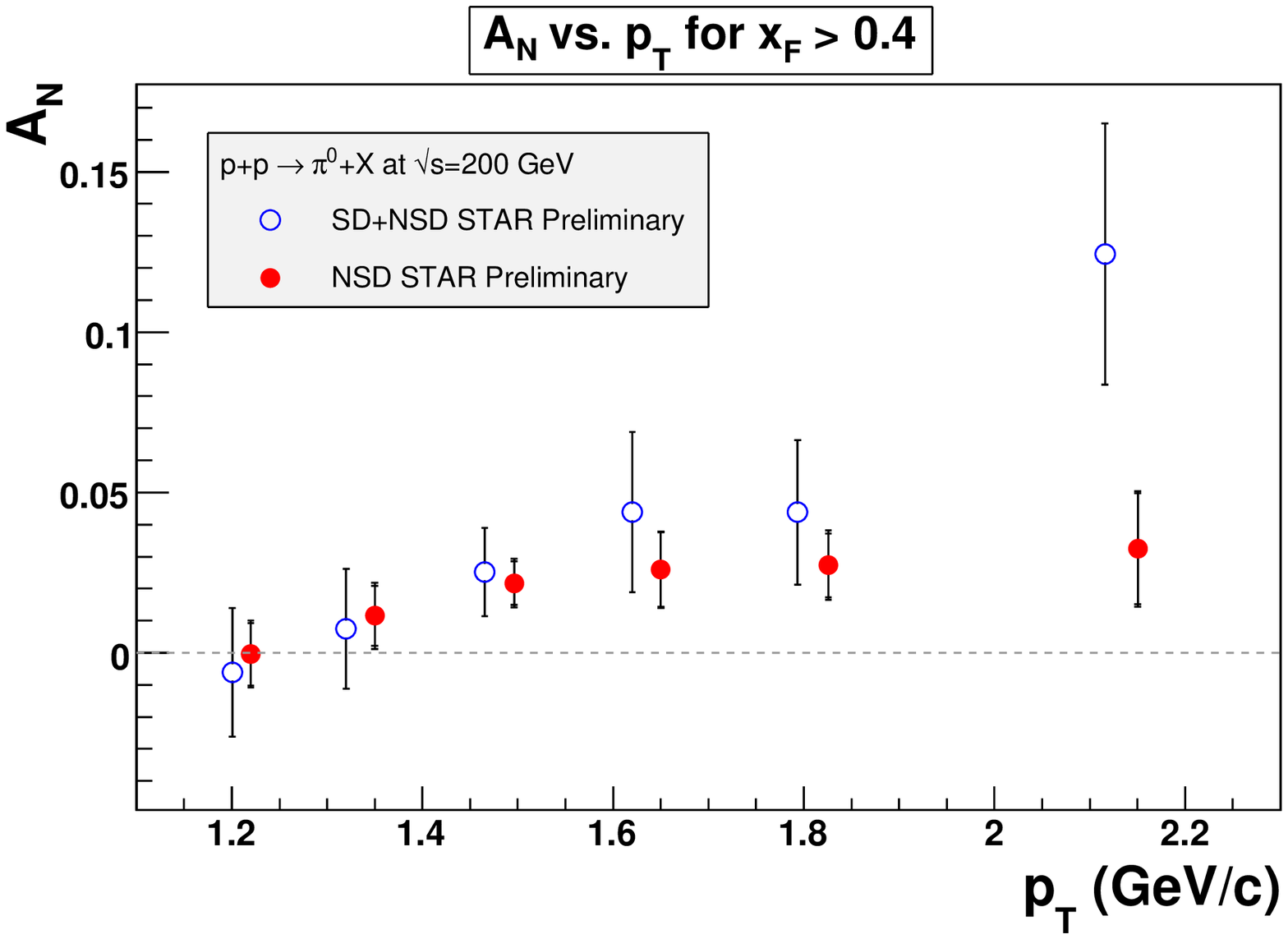}
  \caption{$\pi^{0}$ transverse single-spin asymmetries ($A_{N}$) versus transverse momentum 
  ($p_{T}$). For the left-hand figure and NSD sample (Right) inner error bars are statistical errors only, and outer error bars are statistical plus systematic uncertainties. Points for the mixed event sample (Right) include only statistical uncertainties and have been offset to the left by 0.03 GeV/c.}
   \label{fig:ANvspT}
\end{figure}
Asymmetries shown have been calculated using a cross-ratio method,
\begin{equation}
A_{N}=\frac{1}{P_{beam}}\times\frac{\sqrt{N_{L\uparrow}\times N_{R\downarrow}}-\sqrt{N_{L\downarrow}\times N_{R\uparrow}}}{\sqrt{N_{L\uparrow}\times N_{R\downarrow}}+\sqrt{N_{L\downarrow}\times N_{R\uparrow}}}
\end{equation}
where $N_{L(R)\uparrow(\downarrow)}$ is the number of events in the
left (right) module with spin up (down). Results for $x_{F}<0$ are
consistent with zero, so we show data only for $x_{F}>0$. Results
are shown for two different software conditions on the STAR Beam-Beam Counter (BBC). 
First, a software trigger requiring a coincidence between BBC-east (BBCE) and BBC-west
(BBCW) is required, producing a nearly pure non-singly-diffractive
(NSD) event sample. These data are shown in comparison to previously
published data from RHIC runs 3, 5, and 6 \cite{Abelev:2008}. A second event sample is
analyzed requiring a hit in BBCE, the same side of the collision as the FPD, and no hit in BBCW. 
This results in a mixture of NSD and singly-diffractive (SD) events. Results from analysis of the 
mixed sample are shown in comparison with the nearly-pure NSD event sample. Statistical errors 
are calculated as 
$\Delta A_{N}=1/(P_{beam}\times
\sqrt{N_{L\uparrow}+N_{L\downarrow}+N_{R\uparrow}+N_{R\downarrow}})$
and scaled to account for remaining acceptance asymmetries, which are less than $20\%$. For the
nearly-pure NSD event sample, figures show inner error bars for statistical
errors and outer error bars for systematic errors combined in quadrature
with statistical errors. Errors for the mixed event sample are statistical
only. The systematic uncertainties for the mixed event sample are expected to be less than the 
statistical errors.

\section{Results}

\begin{figure}
  \includegraphics*[scale=0.35]{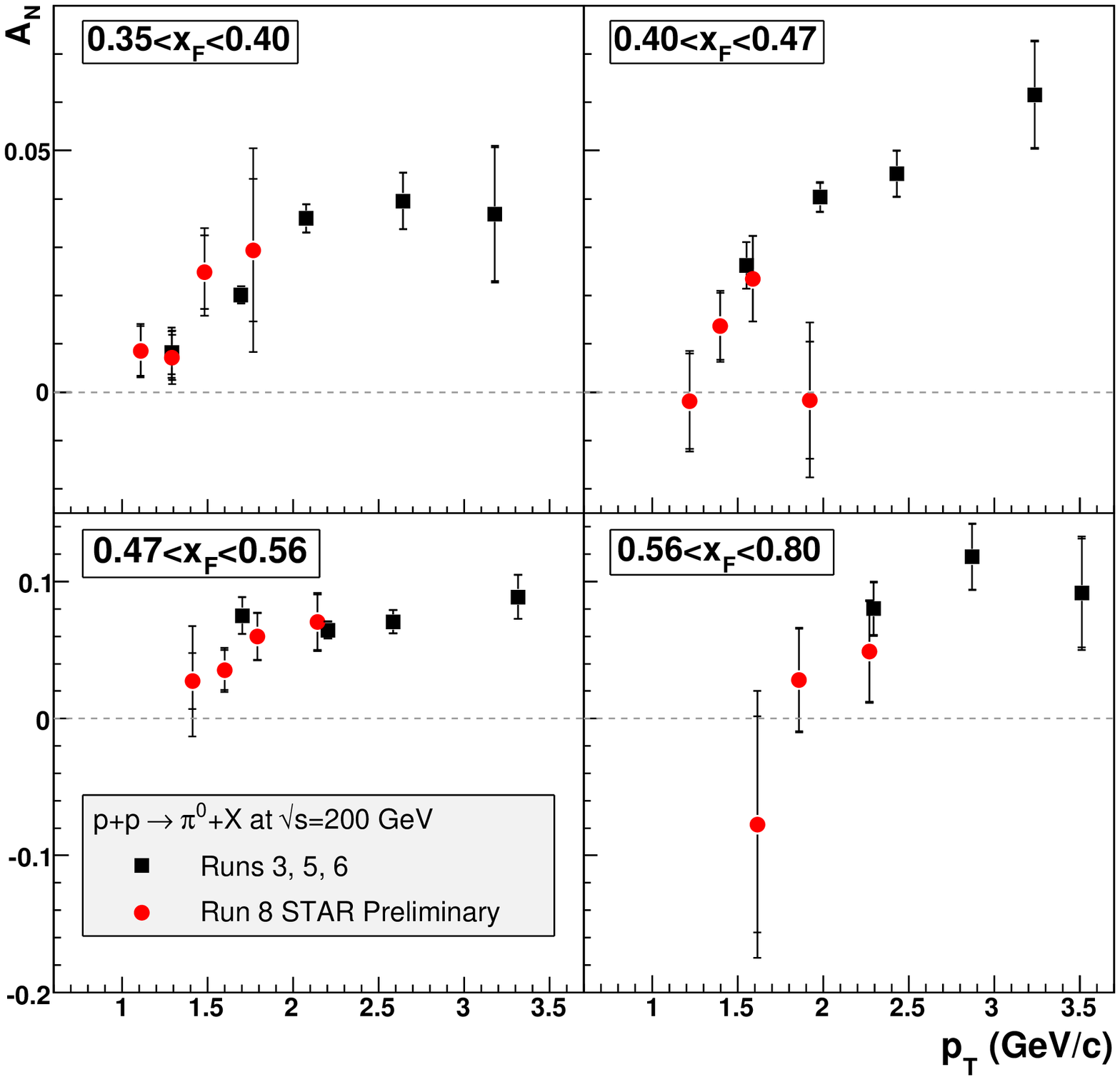}
  \includegraphics*[scale=0.35]{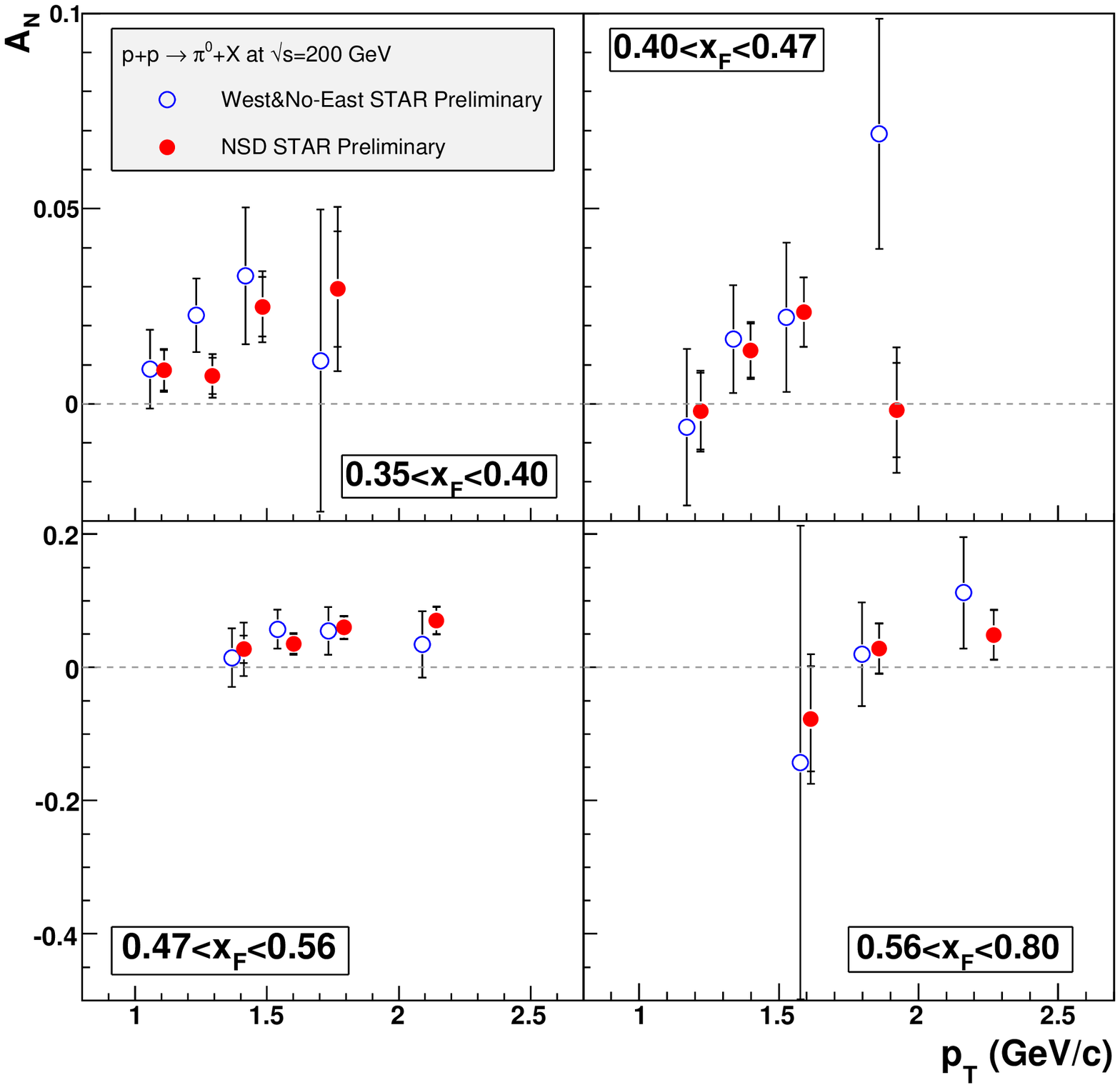}
  \caption{$\pi^{0}$ transverse single-spin asymmetries ($A_{N}$) versus transverse momentum ($p_{T}$) in $x_{F}$ bins. For the left-hand figure and NSD sample (Right) inner error bars are statistical errors only, and outer error bars are statistical plus systematic uncertainties. Points for the mixed event sample (Right) include only statistical uncertainties and have been offset to the left by 0.06 GeV/c.}
  \label{fig:ANvspTandxF}
\end{figure}

Results from RHIC run-8 are mostly consistent with results from previous
runs, Fig.\@ \ref{fig:ANvspT} (Left). However, in the low-$p_{T}$ region where previous
data were dominated by RHIC run 3 with limited statistics, the
run-8 data show $A_{N}$ continuing to decrease. Likewise, observing
the $p_{T}$-dependence at fixed $x_{F}$, Fig.\@ \ref{fig:ANvspTandxF} (Left), run-8 results
show general agreement with previous results in overlapping regions
and smooth fall-off toward lower-$p_{T}$.

The results from the mixed event sample show general agreement with
the nearly-pure NSD sample in terms of $p_{T}$, Figs.\@ \ref{fig:ANvspT} and \ref{fig:ANvspTandxF} (Right).

\section{Summary}

We have shown STAR measurements of transverse single-spin asymmetries,
$A_{N}$, from $p_{\uparrow}+p\rightarrow\pi^{0}+X$ at $\sqrt{s}=200$
GeV from the 2008 RHIC run. The data show general consistency with
previous measurements and show $A_{N}$ continuing to decrease at
low-$p_{T}$. We have also shown results from a mixed event sample
combining singly-diffractive and non-singly-diffractive events. Results
from the mixed event sample are consistent with results from
the nearly-pure non-singly-diffractive event sample. 

%



\bibliographystyle{aipproc}   





\end{document}